\documentclass[fleqn,10pt]{wlscirep}
\usepackage[utf8]{inputenc}
\usepackage[T1]{fontenc}
\usepackage{bm}
\usepackage{lineno}

\title{Imaging hyper-entanglement based on the Hardy-type nonlocality paradox}

\author[1]{Wuhong Zhang}
\author[1]{Xiaodong Qiu}
\author[1]{Dongkai Zhang}
\author[1,*]{Lixiang Chen}

\affil[1]{Department of Physics and Collaborative Innovation Center for Optoelectronic Semiconductors and Efficient Devices, Xiamen University, Xiamen 361005, China}

\affil[*]{e-mail: chenlx@xmu.edu.cn}

\begin{abstract}
The concept of quantum entanglement and hyper-entanglement, lying at the heart of quantum information science and technologies, is physically counter-intuitive and mathematically elusive. We design a polarization-encoded ghost imaging system based on the frame of Hardy nonlocality paradox to visualize the evidence of quantum hyper-entanglement by capturing purely nonlocal photonic events. In two-photon polarization-spatial-mode hyper-entangled state, spatial entanglement conveys the ghost images while polarization entanglement encodes the imaging channels. Then whether imaging the single ghost image of a skull-shape object or not can be a direct yet intuitive signature to support or defy quantum mechanics. We use the contrast-to-noise ratio of ghost images to macroscopically characterize the degree of the violation of locality. We also showcase the nonlocal behavior of violating the locality with a reasonable confidence level of $75\%$, microscopically at the single-pixel level. Our strategy not only sheds new light on the fundamental issue of quantum mechanics, but also holds promise for developing hyper-entanglement-based quantum imaging technology.

\end{abstract}

\begin{document}
\flushbottom
\maketitle

\thispagestyle{empty}

\noindent \textbf{Teaser:} An experimental strategy to visualize the evidence of quantum hyper-entanglement in polarization and spatial mode degrees of freedom is devised

\section*{Introduction}
The ability to capture images with new methods allow us to see more than what can be seen by the unaided eye. Quantum imaging technique is one of the such exciting methods \cite{moreau2019NPR} to understand the quantum world visually. Quantum imaging, also termed ghost imaging, could be traced back to the pioneering work by Shih and co-workers \cite{pitttman1995pra}, who first exploited the entangled photon pairs to image nonlocally. Later, ghost imaging with classically correlated light \cite{bennink2002prl} was also demonstrated, and thus sparking a debate on the quantum versus classical origin of ghost imaging \cite{miles2017ghostimaging}. Current quantum imaging technique is a boarder concept, which exploited quantum correlations to image with several advantages over classical optics. For example, sub-shot-noise imaging can achieve a larger signal-to-noise ratio than classical imaging \cite{lloyd2008enhanced,brida2010experimental}; spatial resolution of the image can be enhanced \cite{tsang2009quantum,shin2011quantum,toninelli2019resolution} even to the Heisenberg limit \cite{unternahrer2018super}; the diffraction limit can be broken \cite{lupo2016ultimate} especially in biological imaging \cite{Taylor2014sub,tenne2019super}. Besides, new ideas are continuously emerging in the field of quantum imaging more recently. Based on induced coherence without induced emission, quantum imaging with undetected photons provides an amazing class of quantum imaging technique that relies on single-photon interference and does not require coincidence detection \cite{lemos2014quantum}. A quantum diffraction imaging technique is proposed to detect the phase information of the matter without interaction \cite{asban2019quantum}. Also, imaging through noise with quantum illumination technique can tolerate much noise and loss towards real-world applications \cite{gregory2020imaging}. However, almost all of the above-mentioned quantum imaging techniques only utilized a single photonic degree of freedom. Hyperentanglement that two photons are simultaneously entangled in more than one degree of freedom, has been proven a promising resource for quantum information processing \cite{deng2017quantum}, such as, for the complete Bell state analysis \cite{walborn2003hyperentanglement}, all-versus-nothing test of quantum nonlocality \cite{cinelli2005all}, direct characterization of quantum dynamics \cite{graham2013hyperentanglement}, quantum
holography without (first-order) coherence \cite{defienne2021polarization}. However, hyper-entanglement-based quantum imaging techniques have not been fully explored.

On the other hand, one recent striking progress was an elegant experiment of imaging the Bell-type nonlocal behavior \cite{moreau2019imaging}. The violation of the Bell's inequality was recognized as the first, paradigmatic test of quantum nonlocality \cite{bell1966problem}. Acquiring the images of such a fundamental quantum effect is not only a demonstration that images can be exploited to access the full range of possibilities allowed in the quantum world, but also opens a new way to show quantum imaging advantages out of quantum-correlated sources that could not be obtained classically. There were also some other early attempts to use the imaging technique to characterize quantum entanglement. Boyer {\it et al.} utilized four-wave mixing in a hot vapor to show that twin images can exhibit localized entanglement \cite{boyer2008entangled}. Robert {\it et al.} demonstrated a real-time imaging of entanglement by registering the spatial mode patterns transferred in a hybrid entangled state \cite{fickler2013real}. Edgar {\it et al.} used an electron-multiplying charge-coupled device (EMCCD) to measure correlations in both position and momentum across a multi-pixel field of view in sparse images \cite{edgar2012imaging}. In these previous schemes, however, post processing of the observed images was needed in order to derive and meet the entanglement criteria.

Here, we propose and demonstrate experimentally a strategy of visualizing quantum hyper-entanglement with a single image, which can be a direct signature to support or defy quantum mechanics, as all the captured photons in such an observed ghost images can be totally nonlocal. In the hyper-entanglement, spatial entanglement is responsible for the formation of ghost image that visualizes the evidence of quantum nonlocality while polarization entanglement encodes the imaging channels in the logical framework of Hardy's nonlocality paradox. In contrast to Bell's inequality, Hardy's theory is an attempt to demonstrate nonlocality without inequalities, which gives a more intuitive way to showcase the quantumntness of the system \cite{hardy1992quantum,hardy1993nonlocality}. Hardy's test has been realized experimentally using two spin-half particles \cite{goldstein1994nonlocality}, or photon pairs entangled in different degrees of freedom such as polarization \cite{torgerson1995experimental}, energy-time \cite{vallone2011testing,fedrizzi2011hardy}, and orbital angular momentum (OAM) \cite{chen2012hardy,karimi2014hardy}. Recently, the Hardy's theory was also extended to n-particle or two-particle system even with a high dimensionality \cite{chen2017experimental,jiang2018generalized,meng2018hardy,zhang2020orbital}. Therefore, it has been routinely considered as a fundamental quantum effect to showcase the existence of quantum nonlocality. Here we further explore the Hardy's theory for realizing the visualization of quantum hyper-entanglement in the form of imaging.
\begin{figure}[t]
\centering
\includegraphics[width=1\columnwidth]{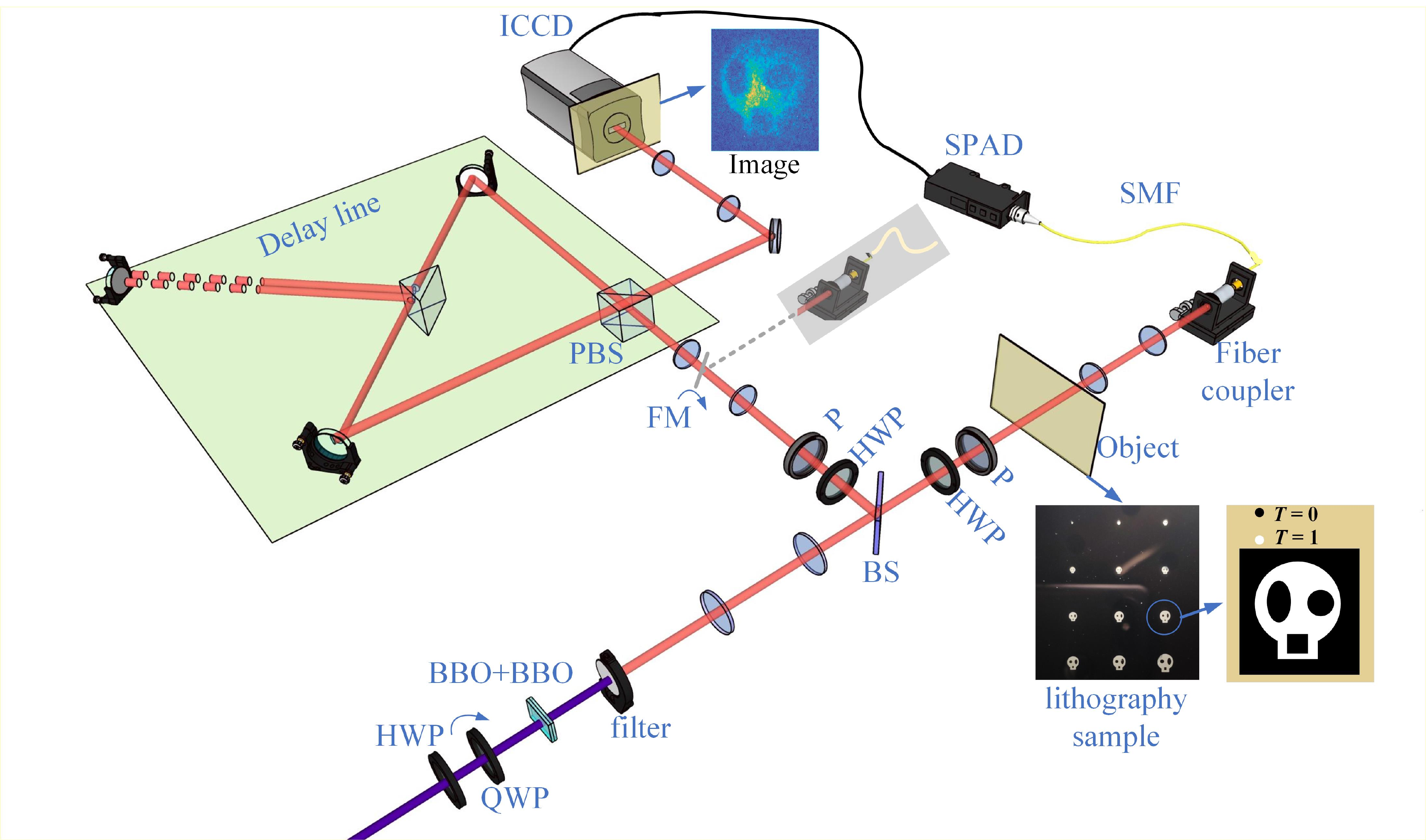}
\caption{Experimental setup for imaging the hyper-entangled state based on the Hardy's nonlocality paradox. HWP: Half wave plate; QWP: Quater wave plate; P: Polarizer; FM: Filp Mirror; PBS: Polarizing beam spliter; SPAD: single-photon detector; SMF: Single Mode Fiber. The object is placed in the signal path while the image is recorded nonlocally with the ICCD camera in the idler path. In each path, the combination of one HWP and one polarizer are used to realize the polarization-encoded ghost imaging channels, which are configured according to the logical structure of Hardy's nonlocality proof. Note that all of the angle of HWP, QWP and P have been calibrated by the coincidence measurement with the photon pairs. This is done by inserting a flip mirror before the delay line at the idler arm to direct the photon to another fiber coupler (shown as gray). An optical delay line for both polarization states $H$ and $V$ are requested in the idler path to compensate the electronic delay in the signal path, which is realized by a modified Sagnac interferometer, see Matherias and Methods for more details.}
\label{fig1}
\end{figure}

\section*{Results}
\subsection*{Theoretical analysis}
As is well known, photon pairs produced via spontaneous parametric down-conversion have many accessible degrees of freedom that can be exploited for the production of entanglement. Here the polarization and transverse spatial modes are simultaneously employed as the hyper-entangled state source. We consider the spatial entanglement in terms of Laguerre-Gaussian (LG) mode \cite{Torres2003quantum}, which can be written as $|\Psi\rangle_{spa}=\sum_{\ell_s,\ell_i,p_s,p_i}{C_{p_s,p_i}^{\ell_s,\ell_i}}|\ell_s,p_s\rangle|\ell_i,p_i\rangle$, where $C_{p_s,p_i}^{\ell_s,\ell_i}$ denotes the amplitude probability of finding one photon in the signal mode $|\ell_s,p_s\rangle$ and the other in the idler mode $|\ell_i,p_i\rangle$, with $\ell$ and $p$ being the azimuthal and radial indices, respectively. Besides, by modulating the polarization state of pump beam, one can neatly generate the polarization entangled state as, $|\Psi\rangle_{pol}= \alpha|H\rangle_s |H\rangle_i + \beta e^{i\phi}|V\rangle_s |V\rangle_i$ \cite{kwiat1999ultrabright,white1999nonmaximally}, where $\alpha$, $\beta$ and $\phi$ characterize the degree of polarization entanglement. Then a two-photon hyper-entangled state can be readily prepared as,
\begin{eqnarray}
 |\Psi\rangle={|\Psi\rangle_{spa}}\otimes{|\Psi\rangle_{pol}}.
\label{eq1}
\end{eqnarray}

\begin{figure}[t]
\centerline{\includegraphics[width=0.6\columnwidth]{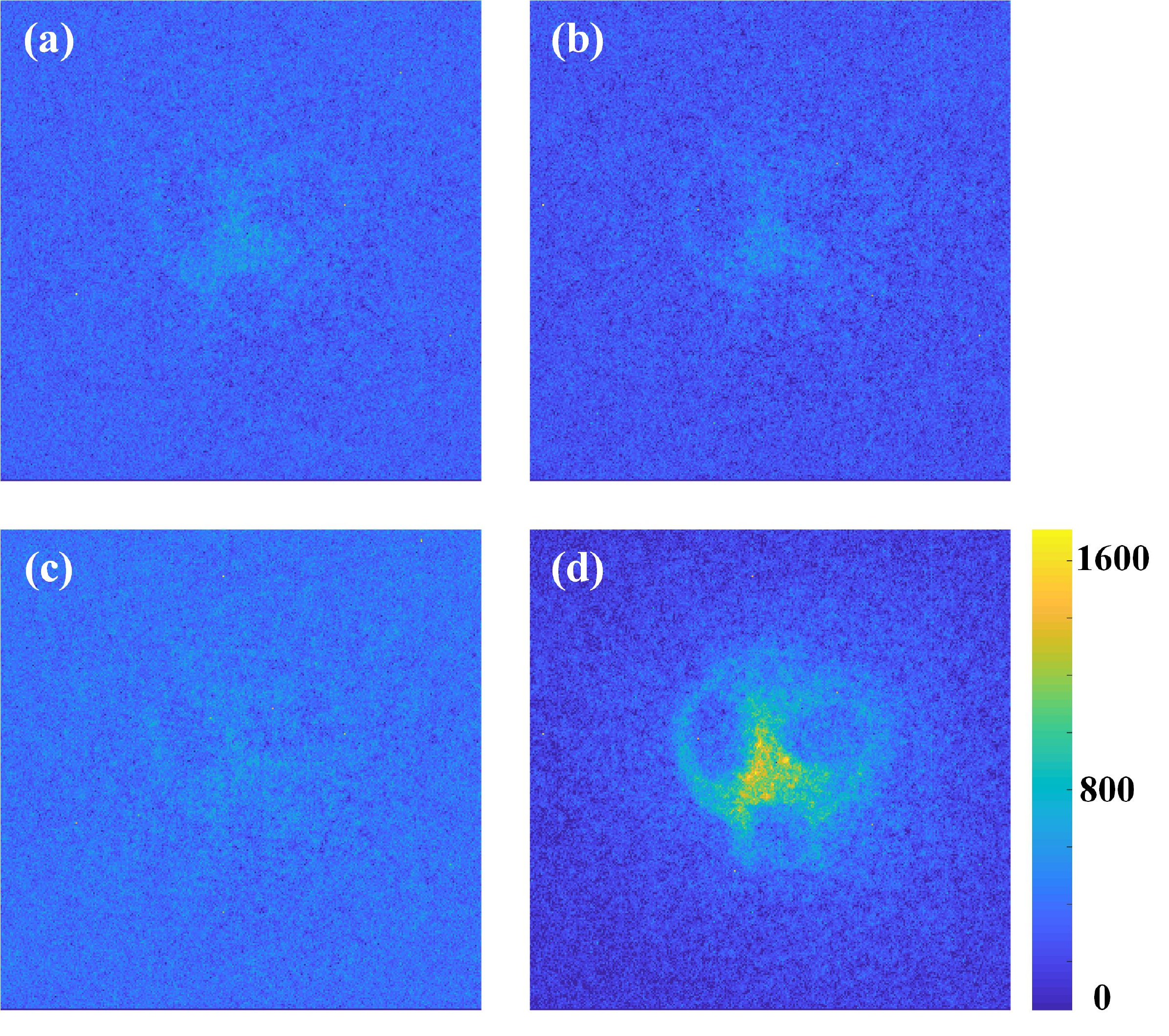}}
\caption{Experimental results for visualizing the Hardy's paradox with an intensity object of skull. The first three ghost images, i.e., (a) $I(\theta_0,\theta_0)$, (b) $I(\bar{\theta}_0,\theta_1)$, and (c) $I(\theta_1,\bar{\theta}_0)$ all have nearly zero intensity. While the fourth one (d) $I(\theta_1,\theta_1)$ reproduces well the image of the skull.}
\label{fig2}
\end{figure}

In order to show the nonlocality directly with the observed images, we exploit the polarization entanglement to encode the imaging channels in the logical proof of Hardy's nonlocality theory, in which the spatial mode entanglement to convey the ghost images. The LG modes form a complete basis in high-dimensional Hilbert space \cite{simon2012two}. Assume any structured object is described by a complex transmission function, $\psi(r,\phi)$, and we denote the photon state described by the same complex wave-funciton as $|\psi\rangle$. Based on the mode expansion method \cite{wuhong2018free,wuhong2019quantum}, one has $|\psi\rangle=\sum_{\ell,p}{{A_{\ell ,p}}{|{\ell,p}\rangle}}$, where $A_{\ell,p}= \langle{\ell,p}|\psi\rangle$ denotes the overlap amplitude. By placing an object with a bucket detector in the path of signal photons, we just project inversely the signal photons into the state, $|\psi\rangle_s=\sum_{\ell_s,p_s}{{A_{\ell_s,p_s}}{|{-\ell_s,p_s}\rangle}}$. Then we can derive the state of idler photons as,
\begin{eqnarray}
|\varphi\rangle_i={_s\langle\psi|} \Psi\rangle_{spa} = \sum_{\ell_s,\ell_i,p_s,p_i}{{A_{\ell_s,p_s}}{C_{p_s,p_i}^{\ell_s,\ell_i}}{|{\ell_i,p_i}\rangle}},
\label{eq2}
\end{eqnarray}
which accounts for the formation of a ghost image of the object. Here, of key interest is the exploitation of the auxiliary polarization degree of freedom to encode the ghost imaging channels within the logical structure of Hardy's nonlocality proof. In the Hardy's paradox, quantum mechanics allows a set of probabilities that is logically inconsistent to that from classical optics. Assume there are two observers measuring dichotomic observables. Alice measures $A_0$ and $A_1$, while Bob measures $B_0$ and $B_1$. $P(A_i,B_j)$ is defined as the joint probability of obtaining $A_i=1$ and $B_j=1$, while $P(\bar{A_i},B_j)$ is that of $A_i=-1$ and $B_j=1$. In the local hidden-variable theory, if the three conditions: (I) $P(A_0,B_0)=0$, (II) $P(\bar{A_0},B_1)=0$ and (III) $P(A_1,\bar{B_0})=0 $ hold, then (IV) $P(A_1,B_1)$ should be exactly zero. However, quantum mechanics allows suitable observables $A_0,B_0$ and $A_1,B_1$, satisfying (I), (II) and (III), but (IV): $P_1=P(A_1,B_1)>0$. It was demonstrated in many literature that the optimal values can reach $P_1=0.09$ \cite{hardy1992quantum,hardy1993nonlocality,goldstein1994nonlocality,torgerson1995experimental}. For this, we assign the polarization angles $\theta_k$ ($k=0, 1$) to the signal and idler photons, as follows,
\begin{equation}
\begin{array}{l}
|A_k\rangle_s=\cos{\theta_k}|H\rangle_s + \sin{\theta_k}|V\rangle_s,\\
|\bar{A}_k\rangle_s=-\sin{\theta_k}|H\rangle_s + \cos{\theta_k}|V\rangle_s,\\
|B_k\rangle_i=\cos{\theta_k}|H\rangle + \sin{\theta_k}|V\rangle_i,\\
|\bar{B}_k\rangle_i=-\sin{\theta_k}|H\rangle_i + \cos{\theta_k}|V\rangle_i.
\label{eq3}
\end{array}
\end{equation}

Due to the mutual independence of polarization and spatial-mode entanglement, we know from Eqs. (1) and (2) that the ghost images sustained by the biphoton polarization channels in the logical structure of Hardy's paradox can be respectively described as,
\begin{eqnarray}
|\varphi\rangle_{i}^{1} &=& {}_s\langle A_0{}|_i\langle B_0|\Psi\rangle_{pol} \otimes {_s\langle\psi|} \Psi\rangle_{spa},\\
|\varphi\rangle_{i}^{2} &=& {}_s\langle \bar{A}_0{}|_i\langle B_1|\Psi\rangle_{pol} \otimes {_s\langle\psi|} \Psi\rangle_{spa},\\
|\varphi\rangle_{i}^{3} &=& {}_s\langle A_1 {}|_i\langle \bar{B}_0 |\Psi\rangle_{pol} \otimes {_s\langle\psi|} \Psi\rangle_{spa},\\
|\varphi\rangle_{i}^{4} &=& {}_s\langle A_1{}|_i\langle B_1|\Psi\rangle_{pol} \otimes {_s\langle\psi|} \Psi\rangle_{spa},
\label{eq4}
\end{eqnarray}

Accordingly, we can observe four ghost images, $I_m = |\langle r, \phi | \varphi\rangle_{i}^{m}|^2$, via four different channels, where $m=1, 2, 3$ and $4$, respectively. In the Hardy's nonlocality proof, we can choose suitable polarization angles to make two-photon joint-detection events via channel $1, 2, 3$ all vanish such that no ghost image can be observed in these channels. Then, whether observing a single ghost image or not via channel $4$ can be a direct yet intuitive signature to support or defy quantum mechanics. Furthermore, since an image is worth a thousand words, each pixel of the image can manifest the nonlocality individually. In other words, our strategy enables the visualization of hyper-entanglement not only with a whole ghost image but also at the single-pixel level via Hardy's nonlocality.

\subsection*{Experimental results}
Our experimental setup is shown in Fig. \ref{fig1}, see Materials and Methods for more details. Firstly, we calibrated our Hyper-entangled source to prepare the non-maximally entangled polarization state. We put a complex object of the skull in the signal path. The skull was fabricated by lithography, which has a size of $1.8mm\times 1.8mm$. As shown by the inset of Fig. 1, it was designed to have a uniform or zero transmission with $T=1$ or $T=0$. Then we record the ghost images with the ICCD camera in the idler path, where four imaging channels are encoded according to Eqs. (4)-(7) with $\theta_1=-18.3^{\circ}$, $\theta_0=34.7^{\circ}$, $\bar{\theta}_1=71.7^\circ$, and $\bar{\theta}_0=124.7^\circ$, respectively. We show our experimental results in Fig. 2, where each ghost image was recorded by accumulating 2500 frames with 5s exposure time. Note that the background and the strong-exposure pixels have been subtracted by evaluating the background with an incorrect gate delay time. From Figs. 2(a), 2(b) and 2(c), we can hardly observe the ghost images of the skull object via the first three channels, $|A_0\rangle_s|B_0\rangle_i$, $|\bar{A}_0\rangle_s|B_1\rangle_i$, and $|A_1\rangle_s|\bar{B}_0\rangle_i$, since all of them lead to zero joint-detection probabilities, $P(A_0, B_0)=P(\bar{A}_0, B_1)=P(A_1, \bar{B}_0)=0$. Therefore, observing a single ghost image from the fourth channel of $|A_1\rangle_s|B_1\rangle_i$ could reveal the quantum feature of nonlocality, in contrast to that local hidden variable theory predicts the disappearance of the image. From Fig. 2(d), we can thus conclude that our experiment confirms the quantum-mechanical prediction while defies the classical theory.

\begin{figure}[t]
\centerline{\includegraphics[width=\columnwidth]{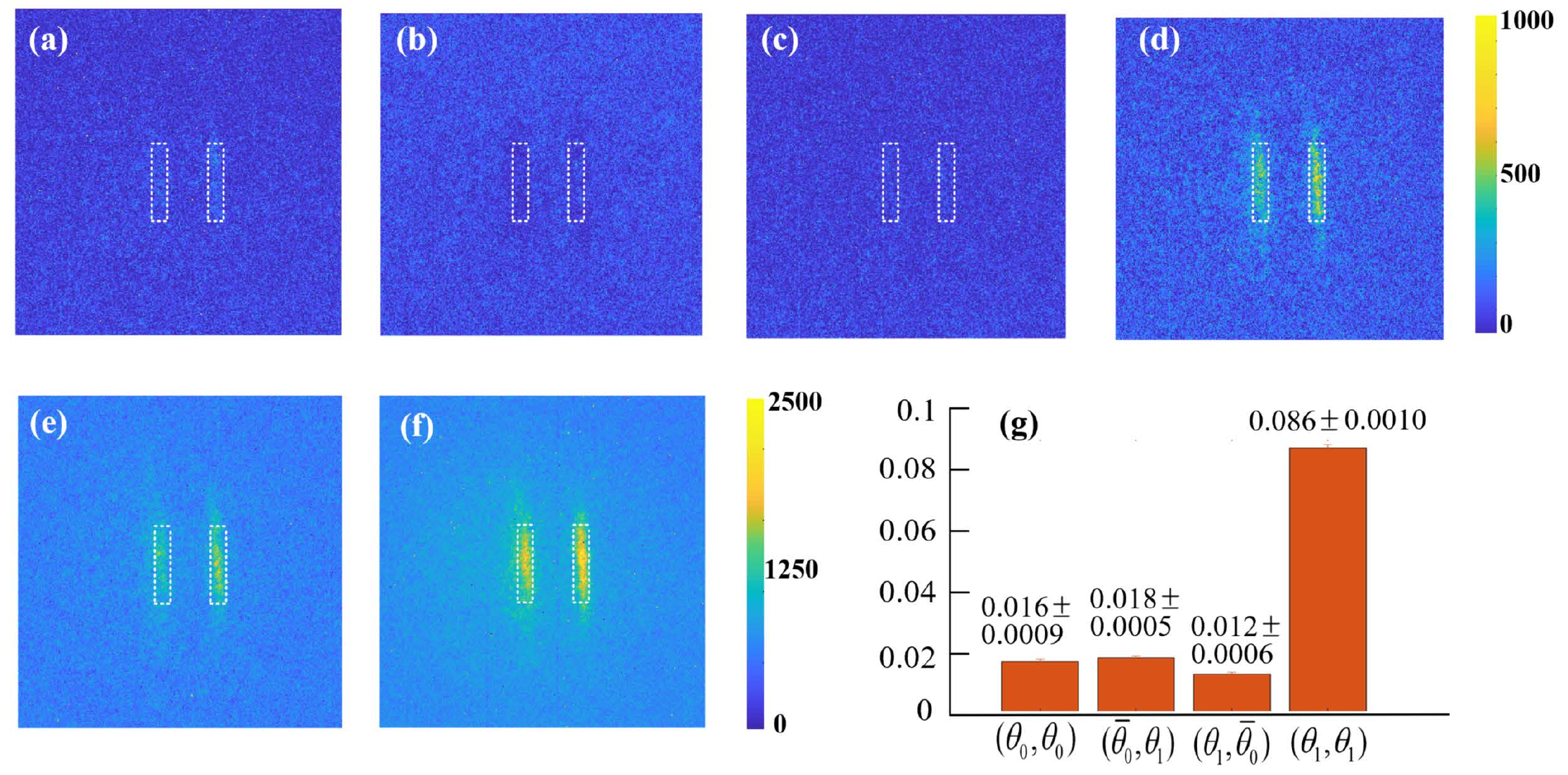}}
\caption{Experimental visualization of Hardy's nonlocality paradox with a double slit. The ghost images: (a) $I(\theta_0,\theta_0)$, (b) $I(\bar{\theta}_0,\theta_1)$, (c) $I(\theta_1,\bar{\theta}_0)$, and (d) $I(\theta_1,\theta_1)$. While (e) and (f) are recorded for $|H\rangle_s|H\rangle_i$ and $|V\rangle_s|V\rangle_i$, respectively. (f) shows the probabilities of photonic events in the Hardy's proof.}
\label{fig3}
\end{figure}

The visibility of these ghost images can also be characterized by using the contrast-to-noise ratio \cite{Chan10OEnoise}, $CNR=(\langle G_{in} \rangle + \langle G_{out} \rangle) /\sqrt{\sigma_{in}^2+\sigma_{out}^2}$, where $\langle G_{in} \rangle$ and $\langle G_{out} \rangle$ are the ensemble average of the photon numbers for any pixel of the skull object with $T=1$ and $T=0$, and $\sigma_{in}$ and $\sigma_{out}$ represent the corresponding variances. Then we can obtain the $CNR=0.08$, $0.07$, $0.05$, and $0.57$ for Figs. 2(a) to 2(d), respectively. Thus, based on hyper-entanglement, we offer an intuitive demonstration of the Hardy's paradox with the single whole ghost images from a macroscopic point of view.

\begin{figure}[t]
\centerline{\includegraphics[width=0.7\columnwidth]{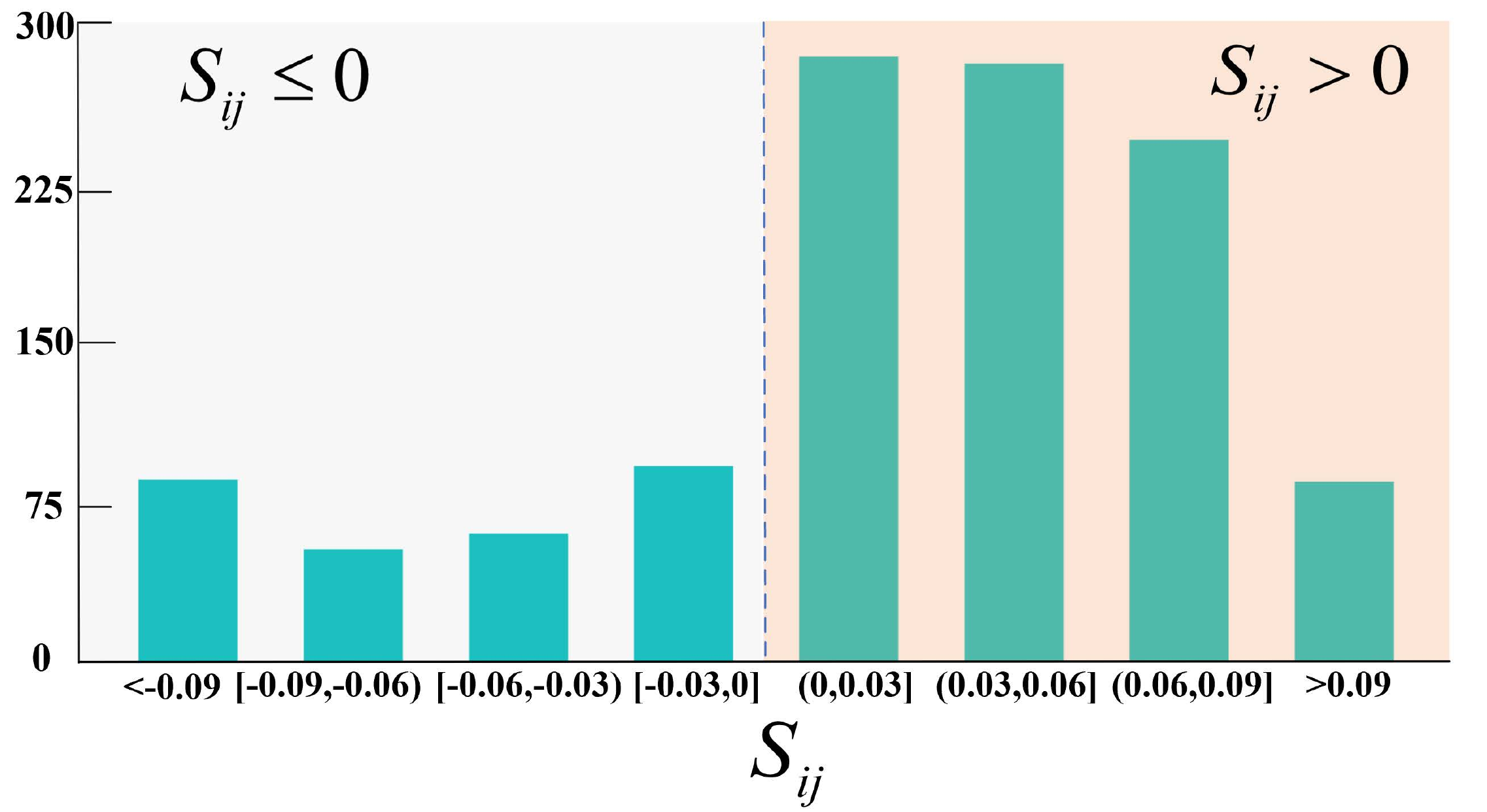}}
\caption{Experimental results of showing the Hardy's paradox at the single-pixel level. The histograms show the statistic analysis of the pixel numbers vs. $S_{ij}$.}
\label{fig4}
\end{figure}

We further exploit a double slit to showcase the Hardy's nonlocality visually even at the single-pixel level. In the same logical structures of Hardy's paradox, we obtain experimentally the ghost images of the double slit, as shown in Fig. 3. Again, the first three ghost images, $I(\theta_0,\theta_0)$, $I(\bar{\theta}_0,\theta_1)$, and $I(\theta_1,\bar{\theta}_0)$, from the polarization-encoded channels remain too difficult to identify. But we can still observe a clear ghost image of the double slit, $I(\theta_1,\theta_1)$, via the fourth channel, $|A_1\rangle_s|B_1\rangle_i$, which is ensured by the quantum-mechanical correlations. We choose the regions of interest (ROI), corresponding to the real double slit, for the single-pixel demonstration of Hardy's paradox. Each ROI of single slit has a width of $275\mu m$, i.e., a pixel-array of $11 \times 51$, as marked by the dotted rectangles in Fig. 3(d). Macroscopically, the total relative photon counts of ROIs for each ghost image can be defined by summing all the pixels' grayscale values within ROI, e.g., $N(\theta_1,\theta_1)=\sum_{i,j} I_{ij}$, where $I_{ij}$ is the grayscale value and $(i,j)$ specifies the pixel's coordinates. To derive the probability of nonlocal events, we also record the images via two orthogonal polarization settings, $|H\rangle_s|H\rangle_i$ and $|V\rangle_s|V\rangle_i$, respectively, see Figs. 3(e) and 3(f). We can then evaluate the total relative number of photon pairs, $N_{total}=N_{ROI}(H,H)+N_{ROI}(V,V)$. Subsequently, we calculate and present in Fig. 3(g) the probabilities for Figs. 3(a)-d(d), respectively. We can see that $P(\theta_1,\theta_1)$ can almost reach the optimal value of $0.09$, while the other three ones remain nearly zero (see Materials and Methods for more details). In an actual experiment, we have to deal with the imperfect states, measurements, and detectors. Thus we adopt the inequality to characterize the degree of violation of localism, namely, $S=P(\theta_1,\theta_1)-P(\bar{\theta}_0,\theta_1)-P(\theta_1,\bar{\theta}_0)-P(\theta_0,\theta_0)=0.0395 >0$. Following the Mermin's theory \cite{mermin1994}, $S>0$ can be seen as the entanglement signature of the system. Thus the obtained $S$ value again show that our obtained ghost images supports the quantum-mechanical prediction offered by the polarization-spatial-mode hyper-entanglement.

Of another particular interest is to further explore the single-pixel information of the images for simultaneous observations of massive nonlocal photonic events. The probability of the nonlocal photonic events for each of the pixels' in ROI can be denoted as $P_{ij}(\theta_1,\theta_1)=I_{ij}/N_{ij}^{total}$. For each pixel, we have the inequality,
\begin{eqnarray}
S_{ij} &=& P_{ij}(\theta_1,\theta_1)-P_{ij}(\bar{\theta}_0,\theta_1)-P_{ij}(\theta_1,\bar{\theta}_0)\\ \notag
  &-& P_{ij}(\theta_0,\theta_0).
\end{eqnarray}

In our experimental results, only those pixels that satisfy $S_{ij}>0$ can be seen as the nonlocal photonic events. To characterize the degree of violation of localism, we make a statistical analysis of $S_{ij}$ within ROI. As illustrated by Fig. \ref{fig4}, most of the pixels have the Hardy's value, $0<S_{ij}<0.09$, in good agreement with the quantum predictions. Specifically, we find that the number of the pixels with $S_{ij}>0$ can reach $N=847$ among the total 1122 pixels of ROI, with a percentage of $75.5\%$. In other words, we have observed the Hardy-type nonlocal events at the single-pixel level with a reasonable confidence level $>75\%$. The other values of $S_{ij}<0$ and even $S_{ij}>0.09$ might be attributed to the imperfect experimental conditions, such as non-uniform illumination, optical misalignment and nonideal optical components.

\subsection*{Conclusion}
In conclusion, we have designed a new strategy of exploring the hyper-entangled features for visualizing the evidence of quantum nonlocality in the Hardy's paradox with ghost imaging system. Based on polarization-encoded imaging channels in the logical structures of Hardy's nonlocality proof, our experimental observations of a single ghost image for a skull object or a double slit can be seen as a direct signature of quantum hyper-entanglement. To characterize the degree of violation of locality, we have demonstrated the Hardy's paradox, macroscopically with the whole ghost images and microscopically at the single-pixel level. In contrast to previous experiments only consider a single degree of freedom, here we exploit the hyper-entanglement for ghost imaging. Both polarization entanglement and spatial mode entanglement are needed to reproduce the desired ghost images. While it is impossible with classical optics only. In other words, our strategy offers an intuitive evidence of Hardy's nonlocality paradox, which, to the best of our knowledge, represents the first attempt to visualize the quantum hyper-entangled state with the ghost imaging technique. Our work also suggests that hyper-entanglement can be explored for shedding new light on the fundamental issue of quantum mechanics and for developing new quantum imaging schemes.

\section*{Materials and Methods}
\subsection*{Experimental setup}
As shown in Fig. \ref{fig1}. A 355nm ultraviolet laser serves as the pump beam, and the spontaneous parametric down-conversion (SPDC) is done in two cascaded BBO crystals of orthogonal fast axes via collinear type-I phase matching. Both crystals are sufficiently thin to ensure the indistinguishability of photon pairs regardless of being emitted from two different crystals. A detailed description of this source was given in \cite{kwiat1999ultrabright,white1999nonmaximally}. One half-wave plate (HWP) together with one quarter-wave plate (QWP) is used to tune the polarization of pump beam such that the desired non-maximally entangled states can be prepared. Then a $50:50$ non-polarizing beam splitter (BS) separates the signal photons from the idler ones. In both paths, we measure the polarization states of idler and signal photons with the combination of one HWP and one polarizer (P), respectively. In the signal path, the object is placed on the image plane of the crystal via a $4f$ imaging system. Another 4f system together with a single-mode fiber (SMF) connected to a single-photon detector (SPAD) serves as the bucket detector to collect single photons, without any spatial resolutions. While in the idler path, we use an intensified CCD (ICCD) camera, also located equivalently on the image plane of the crystal, to record the image. Remember that the ICCD is triggered by the single-photon events conveyed from the single-photon detector in the idler arm, for which an optical delay of about 22 meters is requested to compensate the electronic delay \cite{morris2015imaging}. In our configuration, we built a modified Sagnac interferometer to realize the optical delays for both  $H$ and $V$ polarizations, see \cite{qiu2020parallel} for more details.

\subsection*{Calibration of the Hyper-entangled source }
 Here the polarization and transverse spatial modes are simultaneously employed as our hyper-entangled source. It consists of two adjacent $\beta$-barium borate (BBO) crystals of orthogonal optical axes, both of which are cut for type-I phase matching. Before visualizing the Hardy's paradox with our polarization-encoded ghost imaging system, we use the coincidence circuit to optimize the Hardy's nonlocality proof with polarization entanglement only. For this, a flip mirror before the optical delay line is used to direct the idler photons, instead, to a single-photon detection stage as that in the signal path, and then the coincidence counting is conducted. By carefully tuning the linear polarization of UV pump beam with a half-wave plate (HWP) together with a quarter-wave plate (QWP), we can prepare the optimal non-maximally entangled state as, $|\Psi\rangle_{pol}= \alpha |H\rangle_s |H\rangle_i - \beta |V\rangle_s |V\rangle_i$, with $\alpha=0.43$ and $\beta=0.9$.  In this case, quantum mechanical prediction gives the Hardy's probability $P(A_1, B_1)=(|\alpha|-|\beta|)^2|\alpha\beta|^2/(1-|\alpha\beta|)^2=0.088$, in contrast to $P(A_1, B_1)=0$ from classical optics. It is noted that, our calibration process is the standard way to realize the Hardy's test as implemented in many literature \cite{hardy1993nonlocality,goldstein1994nonlocality,torgerson1995experimental}. Here our calibration process was achieved in several times and tested the stability in 48 hours for imaging consideration.
 \subsection*{Pixels analysis of the obtained ghost image}
 Since the maximum value of $P$ can be only about $0.09$ in the Hardy's theorem, we increased the gain (3800) which is nearly to the maximum (4000) of the ICCD's threshold to have a better contrast of the ghost image. Each of the ghost images was recorded by accumulating 2500 frames with 5s exposure time. The background of each ghost images was evaluated by the same setup but with an incorrect gate delay time. The camera intensifier was triggered for every heralding detection by the SPAD during the exposure time. The ICCD sensor was air cooled to $-30^{\text{o}}{\text{C}}$. The images were thresholded to generate binary images that correspond to the detection of single photons. Each pixels gray value of the image is actually obtained by averaging 2500 times, and each pixels accumulate more than one photon events in a single round exposure time (5s). Thus it is not rigorous to use the number of single photons to do analysis in our high gain mode with long exposure time compared with the previous study \cite{moreau2019imaging}. Instead, here, we define the gray value of each pixels as the relative photon count instead of the real one to analysis the pixel's information. Moreover, in the previous study of analysing the pixel's value of the ghost image obtained by ICCD \cite{moreau2019imaging}, they have divided the many frames into different part to obtain the uncertainties on the mean value. In our test, we found the uncertainties were relatively low since each pixels were averaged electronically and accumulated in many times with the high gain mode and a relatively larger exposure time (five times than ref [22]). So in our final results, we have omitted the error bar of the pixels analysis shown in Figure \ref{fig3}g and \ref{fig4}.
 \subsection*{Calculation of the SNR and S of the ghost image}
\begin{figure}[thb]
\centerline{\includegraphics[width=\columnwidth]{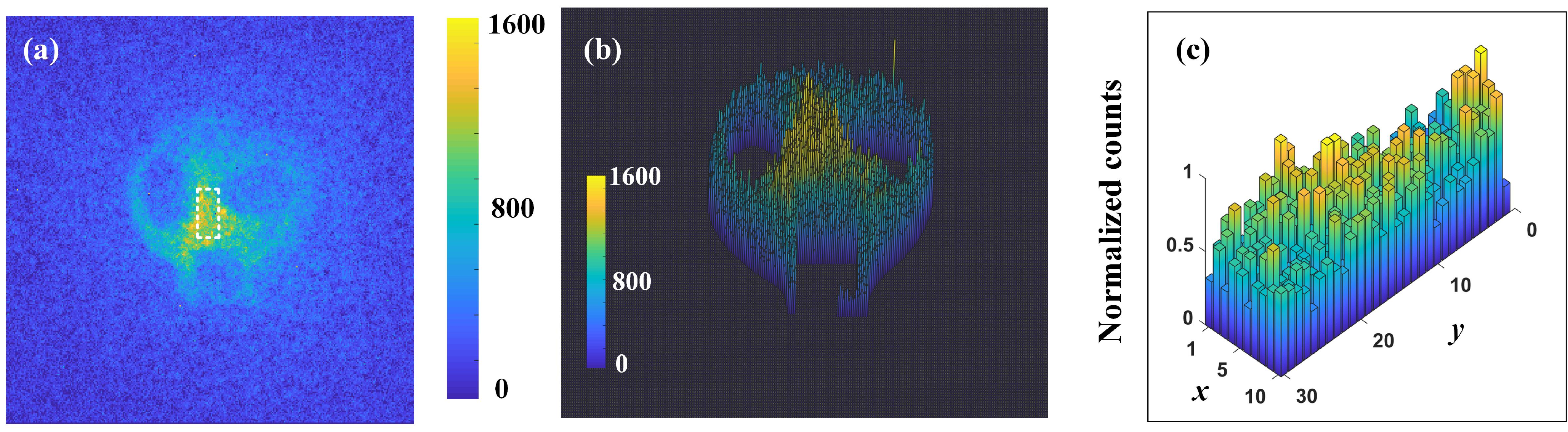}}
\caption{Data processing of the ghost image to obtain the SNR and S. (a) The ghost image obtained from the ICCD; (b) Processed data of the ghost image; (c) Extracted regions of interest from the ghost image. }
\label{fig5}
\end{figure}

The contrast-to-noise ratio \cite{Chan10OEnoise}, $CNR=(\langle G_{in} \rangle + \langle G_{out} \rangle) /\sqrt{\sigma_{in}^2+\sigma_{out}^2}$ is employed to characterize the quality of the obtained ghost image. As can be seen in the Fig. \ref{fig5}, the ghost image of $I(\theta_1,\theta_1)$ is shown in the Fig. \ref{fig5} (a). Then we use the designed object to cutting up the interested area, as plotted by 3-D pixels relative photon count in the Fig. \ref{fig5} (b). The value of the relative photon count can be used to calculate the $\langle G_{in} \rangle$. While the rest of the ghost image can be seen as $\langle G_{out}\rangle$. A ROI ($10*30$ pixels) of the ghost image is picked up as a white dot rectangular plotted in Fig.\ref{fig5} (a). The normalized relative photon count of each pixels in the ROI is shown in Fig.\ref{fig5} (c), and $x,y$ denote the pixels position. As can be intuitively seen from the Fig.\ref{fig5} (c), the value of each pixels can be obtained for further calculation. Thus we can obtained the CNR of each ghost images, as well as the S and $S_{ij}$ in Eq.(8).

\bibliography{myref}

\section*{Acknowledgements}
This work is supported by the National Natural Science Foundation of China (12034016, 61975169 and 11904303), the Fundamental Research Funds for the Central Universities at Xiamen University (20720190054, 20720190057,20720200074), the Natural Science Foundation of Fujian Province of China for Distinguished Young Scientists (2015J06002), the Youth Innovation Fund of Xiamen (3502Z20206045) and the program for New Century Excellent Talents in University of China (NCET-13-0495).

\section*{Author contributions}
L.C., W.Z. and X.Q. conceived and designed the experiment. W.Z. conducted the experiment with the help from X.Q. and D.Z. W.Z. and L.C. analysed and interpreted the results. All authors contributed to the writing of the manuscript.

\section*{Competing interests}
The authors declare no competing interests.

\end{document}